\documentclass[aps,prl,twocolumn,showpacs]{revtex4}
\usepackage{graphicx,amsmath,amssymb}
\usepackage{hyperref}
\usepackage{color}

\begin{document}

\title{Jastrow form of the Ground State Wave Functions for Fractional Quantum Hall States}	
\author{Sutirtha Mukherjee$^{1}$ and Sudhansu S Mandal$^{1,2}$}
\affiliation{$^{1}$Department of Theoretical Physics, Indian Association for the Cultivation of Science, Jadavpur, Kolkata 700032, India}
\affiliation{$^{2}$Department of Physics and Centre for Theoretical Studies, Indian Institute of Technology, Kharagpur 721302, India }
\date{\today}
\begin{abstract}
	The topological morphology--order of zeros at the positions of electrons with respect to a specific electron--of Laughlin state at filling fractions $1/m$ ($m$ odd) is homogeneous as every electron feels zeros of order $m$ at the positions of other electrons.
	Although fairly accurate ground state wave functions for most of the other quantum Hall states in the lowest Landau level are quite well-known, it had been an open problem in expressing the ground state wave functions in terms of flux-attachment to particles, {\em a la}, this morphology of Laughlin state.  With a very general consideration of flux-particle relations only, in spherical geometry,
	we here report a novel method for determining morphologies of these states. Based on these, we construct 
	almost exact ground state wave-functions for the Coulomb interaction. 
	Although the form of interaction may change the ground state wave-function, the same morphology constructs the latter irrespective of the nature of the interaction between electrons.
\end{abstract}	
\maketitle

Strong correlation between electrons in the two dimensional electron systems subjected to perpendicular magnetic field creates non-trivial topological orders \cite{Wen95} that are manifested  through distinct fractional quantum Hall effect (FQHE) states \cite{Tsui82}. 
 The state at $\nu = 1/m$ ($m$ odd) is very accurately described by the Laughlin wave function \cite{Laughlin83} that has unique and uniform topological structure, {\it i.e}, every electron feels zeros of order $m$ at the positions of all other electrons. Similar microscopic topological structures for other FQHE states  \cite{Jain89,Mukherjee12,Mukherjee13,Mukherjee15B} in the lowest Landau level
 have not been achieved, due to complex form of the wave functions.
 We here aim to reveal the morphologies that define topological structures, of most of these states. We show that the morhology of a state is as fundamental as its ground state wave function.

 If a fractional quantum Hall effect state occurs at the number of flux quanta $ N_\phi$ for a system of $N$ electrons, the corresponding wavefunction (ignoring ubiquitous Gaussian factors in disk geometry) will be a polynomial of order $N_\phi$ with respect to any electronic coordinate $\xi_j=(x_j-i y_j)$ and the sum of the exponents on all coordinates in each term of the polynomial will be the total angular momentum $M = NN_\phi /2$. As suggested \cite{Laughlin83} by Laughlin, the polynomial in such a case may be expressed in terms of a Jastrow form \cite{Jastrow_PR55} $f(z_{ij})$ with $z_{ij}=\xi_i-\xi_j$, and the exponent $m$ of any $z_{ij}$ determines the order of zeros at the position of $j$-th electron felt by $i$-th electron and {\it vice versa}. 
In a Laughlin wavefunction, the exponents are same for all $z_{ij}$ $(i<j)$ and this homogeneous exponent $(m)$ describes filling factor $\nu = 1/m$. We thus define ``morphology" of this state as ${\cal M}_{1/m}^{(N)} =\left[ \lambda_1,\lambda_2,\cdots , \lambda_{N-1} \right]$ where $\lambda_k = m$ represents order of zeros  felt by any electron at the positions of other electrons and suffix $k$ labels $(N-1)$ zeros for any specific electron. This morphology that provides the relation between flux and particles, $N_{\phi} = m(N-1)$ is derived from the wavefunction. In the following paragraph, we demonstrate an inverse approach: morphology is obtained from the flux-particle relation (FPR) at which the incompressible ground state occurs in a spherical geometry \cite{Haldane83}, and followed by the determination of the wave function for the ground state.

The FQHE state at $\nu = 1/3$ corresponds \cite{Fano_PRB86} to $N_\phi = 3N-3$. If $N=2$, $ N_\phi =3 $ and hence we write corresponding morphology as $ {\cal M}_{1/3}^{(2)}=\left[ 3_{1} \right]$ which immediately suggests the wavefunction $\Psi_{1/3}^{(2)}=z_{12}^3$.  If $N=3$, $ N_\phi =6 $ and hence we write corresponding morphology as $ {\cal M}_{1/3}^{(3)}=\left[ 3_{1},3_{2} \right]$ that suggests the wavefunction $\Psi_{1/3}^{(3)}=z_{12}^3z_{13}^3z_{23}^3$. In general,
$N_\phi = 3N-3$ corresponds to the morphology ${{\cal M}_{1/3}^{(N)}
=[3_1,3_2,\cdots , 3_{N-1}]}$ and hence we obtain Laughlin wavefunction \cite{Laughlin83} $\Psi_{1/3} = \prod_{i<j}^{N}z_{ij}^3$ which is inevitable for this particular FPR. In this letter, we demonstrate that this principle determines morphology and hence ground state wave function of any FQHE state with filling factors $\nu = n/(2sn \pm 1)$, where $n$ and $s$ are integers.

\begin{figure}
	\includegraphics[height=2.5cm]{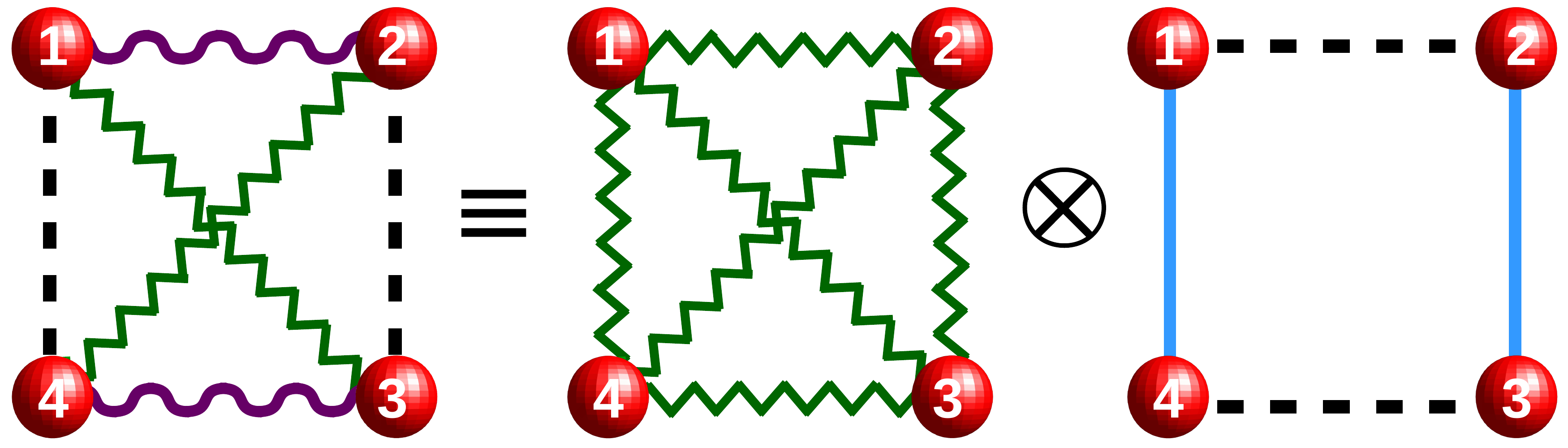}
	\caption{Graph for $N=4$ at $\nu = 2/5$. The filled circles  
		with associated numbers represent different electrons at the vertices of the graphs.
		Links between vertices represented by dashed, solid, sinusoidal, and wiggly lines 
		correspond to exponents $1,\, -1,\, 3$, and 2 respectively on the link variable $z_{ij}$. 
		The graph at the left panel representing ${\cal M}_{2/5}^{(4)} =[1_1,2_2,3_3]$ is equivalent to the direct product of ${\cal M}_{L,1/2}^{(4)}=[2_1,2_2,2_3]$ (graph at central panel) and $\bar{{\cal M}}_{2/5}^{(4)}=[(-1)_1,0_2,1_3]$ (graph at right panel).     }
	\label{Fig1}
\end{figure}

The FQHE state at $2/5$ occurs for the FPR \cite{He92}, $N_\phi = (5/2)N-4$ and so $N$ must be even. For $N=2$, $N_\phi = 1$ and hence ${\cal M}_{2/5}^{(2)} = [1_1]$. If $N$ increases by 2, $N_\phi$ increases by 5. The additional five zeros will be 
shared by the positions of two new particles in two possible ways: 4 and 1 or 3 and 2. However, the former possibility is ruled out as 2/5 state is obtained by reducing flux from 1/3 state in which exactly three zeros are situated at the positions of the particles. Therefore, the corresponding morphology for $N=4$ is given by 
${\cal M}_{2/5}^{(4)} = \left[ 1_1,2_2,3_3 \right]$ (Fig.\ref{Fig1}). Reducing 2 from each of the elements in ${\cal M}_{2/5}^{(4)}$, {\it i.e.}, factoring out the morphology 
${\cal M}_{L,1/2}^{(4)} = [2_1,2_2,2_3]$ corresponding to bosonic Laughlin state at half filling
represented by the Jastrow factor $\prod_{i<j}^{4} z_{ij}^2$, we find the reduced morphology as $\bar{{\cal M}}_{2/5}^{(4)} = \left[(-1)_1, 0_2, 1_3\right]$, as shown
in Fig.\ref{Fig1}.
$\bar{{\cal M}}_{2/5}^{(4)} = \left[(-1)_1, 0_2, 1_3\right]$ provides the functional form $z_{12}z_{34}z_{23}^{-1}z_{14}^{-1}$. Since the electrons are indistinguishable, 
all possible permutations of the positions of electrons at the vertices of Fig.\ref{Fig1} need to be considered. With all six possible permutations, we obtain the wave-function for 4-particle 2/5 state as
\begin{eqnarray}
&& \Psi_{2/5}^{(4)} = \prod_{i<j}^4 z_{ij}^2 [ z_{12}z_{34}\left( z_{13}^{-1}z_{24}^{-1} + z_{14}^{-1}z_{23}^{-1} \right) 
+ z_{13}z_{24} \nonumber \\
&    \times & \left( z_{14}^{-1}z_{23}^{-1} - z_{12}^{-1}z_{34}^{-1} \right) - z_{14}z_{23} \left( 
z_{12}^{-1}z_{34}^{-1} + z_{13}^{-1}z_{24}^{-1} \right) ]\quad
\label{Eq1}
\end{eqnarray}
Since only one graph (see Fig.\ref{Fig1}) is possible for $\bar{{\cal M}}_{2/5}^{(4)}$, 
the corresponding wave function should be the only possible wave function and therefore must be the exact ground state wave function, no matter what the form of 
interaction is as long as
 it produces an incompressible state. Indeed,
in a spherical geometry \cite{Wu76} with $z_{ij} \equiv u_iv_j -v_iu_j$ where spherical spinors $u_j = \cos (\theta_j /2) e^{i\phi_j/2}$ and $v_j = 
\sin (\theta_j /2) e^{-i\phi_j/2}$ for the spherical co-ordinate systems $ 0 \leq \theta_j \leq \pi$ and $0 \leq \phi_j \leq 2\pi$, $\Psi_{2/5}^{(4)}$ is 
found to be the {\em exact} Coulomb ground state.

The indistinguishable nature of electrons demands addition of all these six terms with
appropriate sign due to fermionic nature of electrons for constructing 
antisymmetric many body wave function (\ref{Eq1}). The individual zeros of all these terms 
are at the positions
of the particles. But all the zeros of $\Psi_{2/5}^{(4)}$ are not at the positions of the electrons: When all but one electron coordinate, say
$\xi_1$, are kept fixed,  the zeros are at $ \xi_2,\,\xi_3, \,\xi_4,\, (\xi_3z_{24}+\xi_4z_{23})/(z_{23}+z_{24}),\, (\xi_3z_{24}+\xi_2z_{34})/(z_{24}+z_{34}),\, 
(\xi_4z_{23}-\xi_2z_{34})/(z_{23}-z_{34})$; three zeros of first order are at the positions of three electrons and the other zeros have moved from the particle positions. This shifting of zeros from particle positions is consistent with previous numerical study \cite{Graham03} using composite-fermion wave function in a spherical geometry.
Note that each of the six terms in Eq.~(\ref{Eq1}) corresponding to six permutations has identical topological structure, {\it i.e.,} all of these interpret that each electron feels zeros of one third order, 
one second order, and one first order at the positions  of other three electrons. However, the order of zeros at the position of an electron felt by other electrons
may vary from one through three. Although mutually felt order of zeros between two electrons is same in each term, these may vary between these six terms.

If we further increase two electrons, the additional five vortices will be accommodated in the morphology by increasing two elements of which one will be 2 and the other will be 3, {\it i.e.}, ${\cal M}_{2/5}^{(6)} = [1_1,2_2,2_3,3_4,3_5]$.
After reduction of 2 from each of the elements, the reduced morphology will be  $\bar{{\cal M}}_{2/5}^{(6)} = [(-1)_1,0_2,0_3,1_4,1_5]$. With this $\bar{{\cal M}}_{2/5}^{(6)}$, We can draw three topologically distinct graphs (unlike $\bar{{\cal M}}_{2/5}^{(4)}$ where only one graph is possible) which are shown in Table-\ref{Table1}. Each of these graphs provides an anti-symmetric function when all the possible permutation of the electrons at the vertices of the graphs are taken into account. Out of these three possible functions, only two are linearly independent and thus  
a linear combination of these two independent functions will describe the ground state wave function. Indeed, a linear combination of these two functions has more than 99.97$\%$ overlap with the exact Coulomb ground-state. The number of topologically distinct graphs increases with $N$, but for convenience, we consider only a few suitable graphs yet producing very high overlap \cite{Note} with the exact Coulomb ground state when we make a linear combination of the antisymmetric wave-functions constructed from each of them. The similar details for $N=8$ and $10$ are tabulated in Table-\ref{Table1}.       

\begin{table}[htb]
	\caption{The reduced morphology $\bar{{\cal M}}_{2/5}^{(N)}$ after factoring out ${\cal M}_{L,1/2}^{(N)}$ from
		original morphology ${\cal M}_{2/5}^{(N)}$ at $\nu = 2/5$ for different $N$ and $2Q$. The graphs are based on 
		$\bar{{\cal M}}_{2/5}^{(N)}$ where filled circles with associated numbers represent different electrons and each of the dotted(solid) line in the graphs represents exponent $1(-1)$ on the link variable $z_{ij}$ between vertices $i$ and $j$.
		 The Roman numerals label the graphs and the numbers at the bottom of each graph 
		 represents relative weight factors in their linear combination forming a wave function $\Psi_{2/5}^{N}$, which enable us
		to obtain maximum overlap ${\cal O}$ of $\Psi_{2/5}^{N}$ with exact Coulomb ground state. The numbers within parenthesis of ${\cal O}$ represents Monte Carlo uncertainty in the last digit. The functions obtained from the right-most graph for $N=6$ is not considered as it is not linearly indepnedent.   
	}
	\label{Table1}        
	\begin{tabular}{ c | c |c| c |  c  } \hline\hline
		& & & &  \\
		$\,\,N\,\,$& $\,\,2Q\,\,$ & $\bar{{\cal M}}_{2/5}^{(N)} $ & Graphs  &$\,\,\mathcal{O}\,\,$  \\ 
		& & & &  \\ 
		\hline \hline
		6  & 11 &
		\rotatebox[]{90}{\parbox[h]{2cm} {\scriptsize$\big[(-1)_1,0_2,0_3,$\\$1_4,1_5\big]$}}    &
		\begin{minipage}{.30\textwidth}
			\includegraphics[width=40mm]{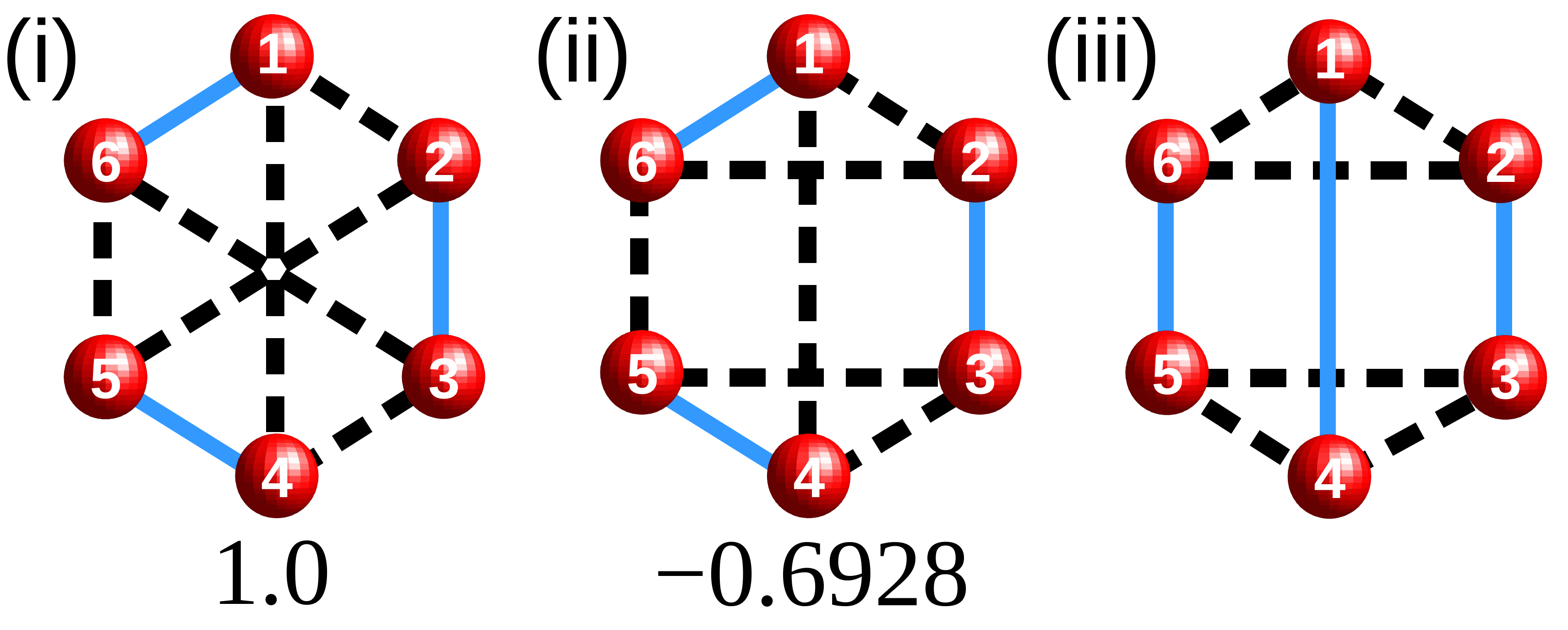}
		\end{minipage} 	 
		& \rotatebox[]{90}{0.9998(1)} \\\hline
		& & & &  \\
		8  & 16 & 
		\rotatebox[]{90}{\parbox[h]{2cm} {\scriptsize$\big[(-1)_1,0_2,0_3,0_4,$\\$1_5,1_6,1_7\big]$}}    & \begin{minipage}{.30\textwidth}
			\includegraphics[width=45mm]{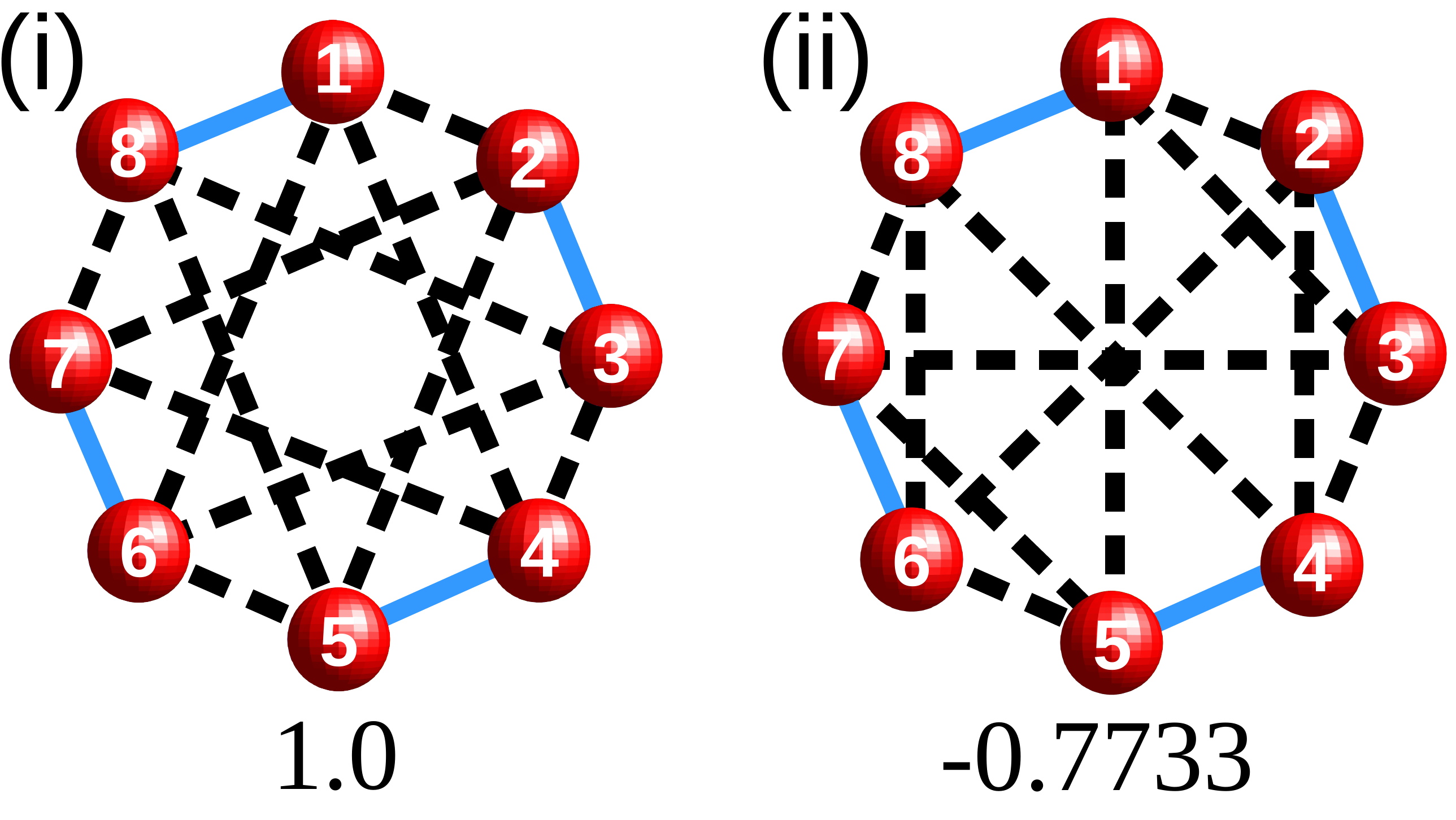} \end{minipage}     
		& \rotatebox[]{90}{0.9997(1)} \\  \hline
		& & & & \\
		10  & 21 &  
		\rotatebox[]{90}{\parbox[h]{4cm} {\scriptsize$\big[(-1)_1,0_2,0_3,0_4,0_5,1_6,1_7,1_8,1_9\big]$}}    & \begin{minipage}{.30\textwidth}
			\includegraphics[width=52mm]{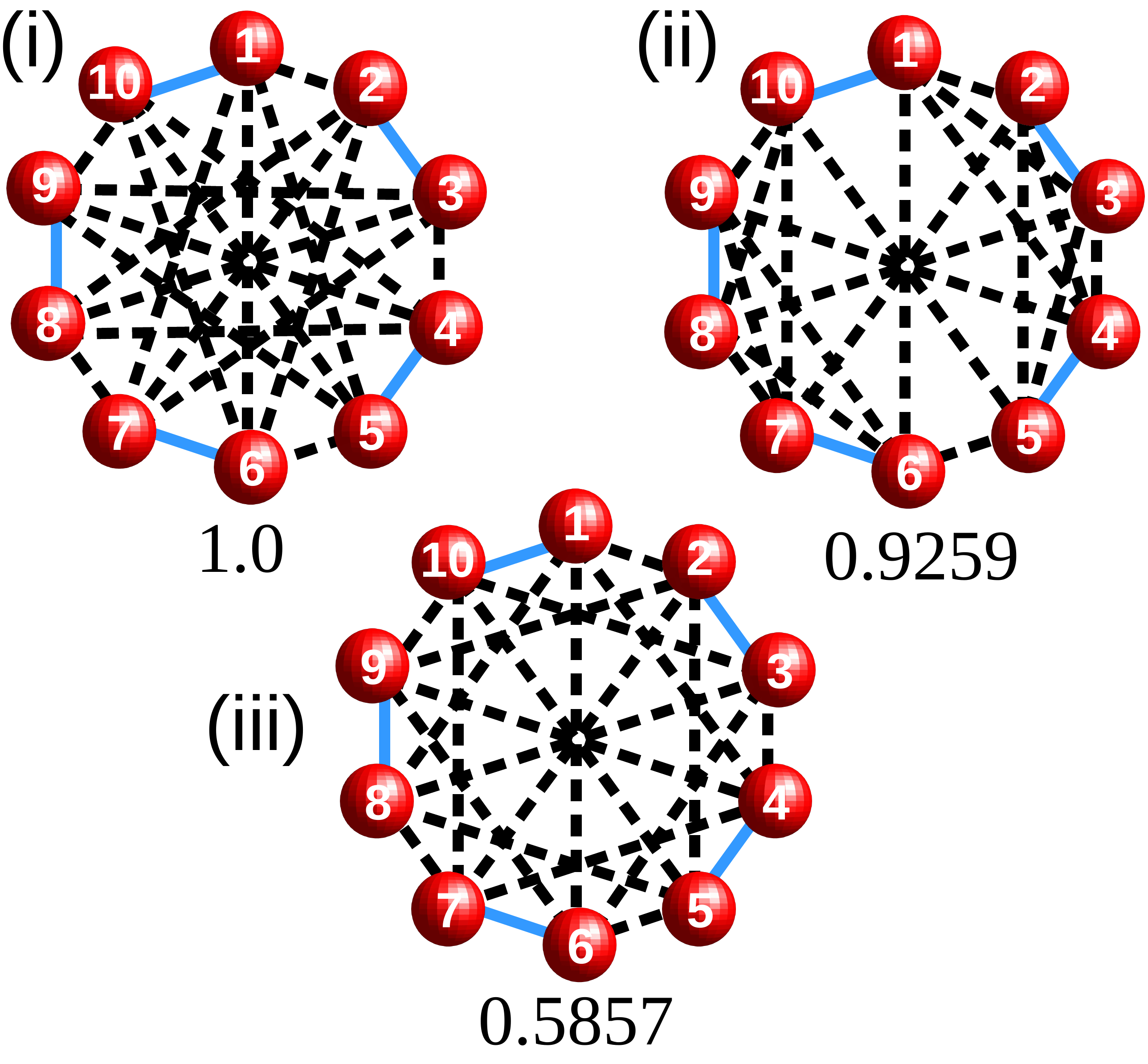}\end{minipage} 
		&\rotatebox[]{90}{ 0.9982(1)} \\
		\hline\hline
	\end{tabular}
\end{table}

\begin{table}[htb]
	\caption{Same as described for $\nu=2/5$ in Table-\ref{Table1} replacing $\nu$ by $2/3$. 
	}
	\label{Table2}        
	\begin{tabular}{ c | c |c| c | c  } \hline\hline
		& & & & \\
		$\,\,N\,\,$& $\,\,2Q\,\,$ & $\bar{{\cal M }}_{2/3}^{(N)}$ & Graphs  &{$\,\,\mathcal{O}\,\,$} \\
		& & & & \\ \hline\hline
		6  & 9  &\rotatebox[]{90}{\parbox[t]{1.4cm} { \scriptsize$\big[1_1,0_2,0_3,$\\$(-1)_4,(-1)_5\big]$}} &
		 \begin{minipage}{.30\textwidth}
		\includegraphics[width=40mm]{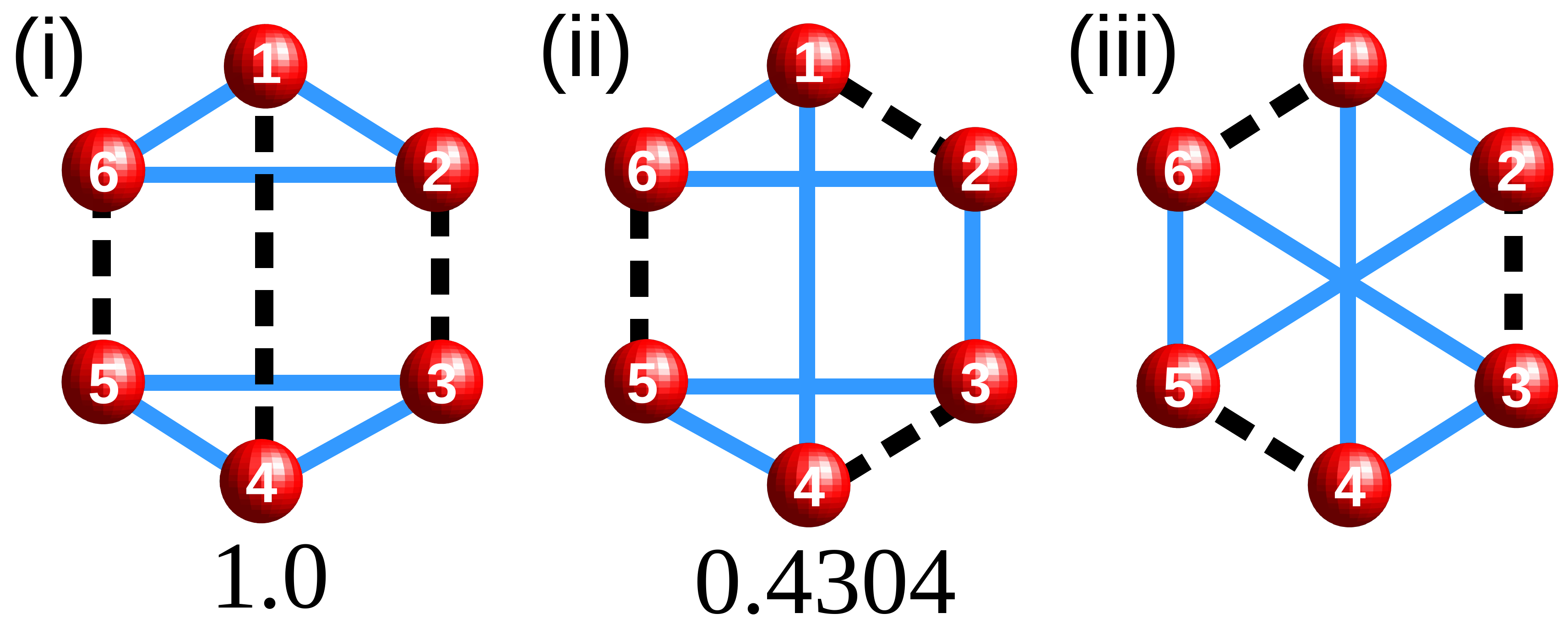}
		\end{minipage} 	
		&\rotatebox[]{90}{ 0.9999(0)} \\\hline
		8  & 12  & \rotatebox[]{90}{\parbox[h]{2.8cm}{ \scriptsize$\big[1_1,0_2,0_3,0_4,(-1)_5,\,$\\$(-1)_6,(-1)_7\big]$}}    & 
		\begin{minipage}{.30\textwidth}
		\includegraphics[width=45mm]{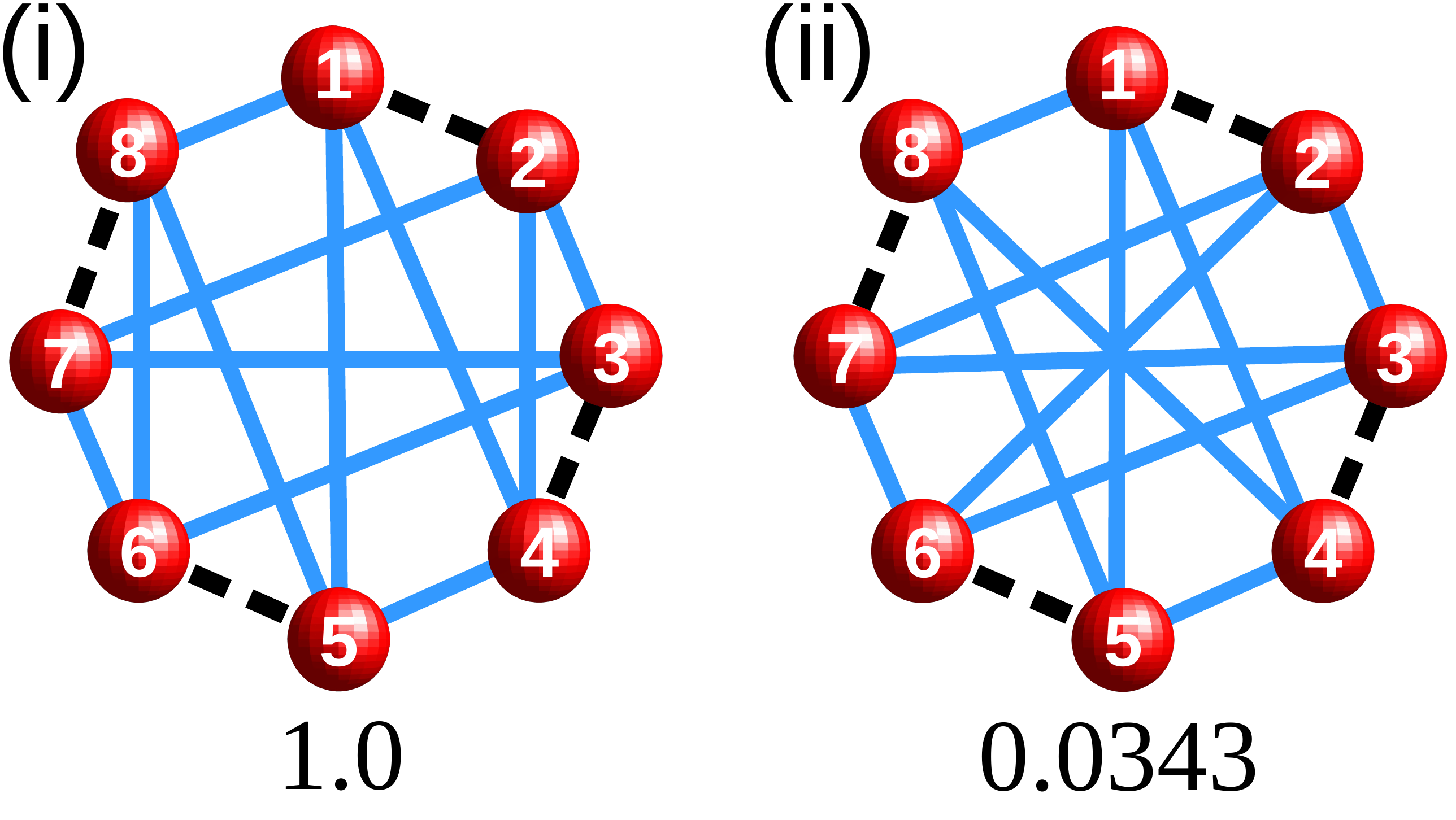} \end{minipage}  
		&\rotatebox[]{90}{ 0.9999(0)} \\ \hline
		10  & 15  &  \rotatebox[]{90}{\parbox[h]{3cm}{\scriptsize$\big[1_1,0_2,0_3,0_4,0_5,$\\$(-1)_6,(-1)_7,(-1)_8,(-1)_9\big]$}}   
		 & \begin{minipage}{.30\textwidth}
		\includegraphics[width=52mm]{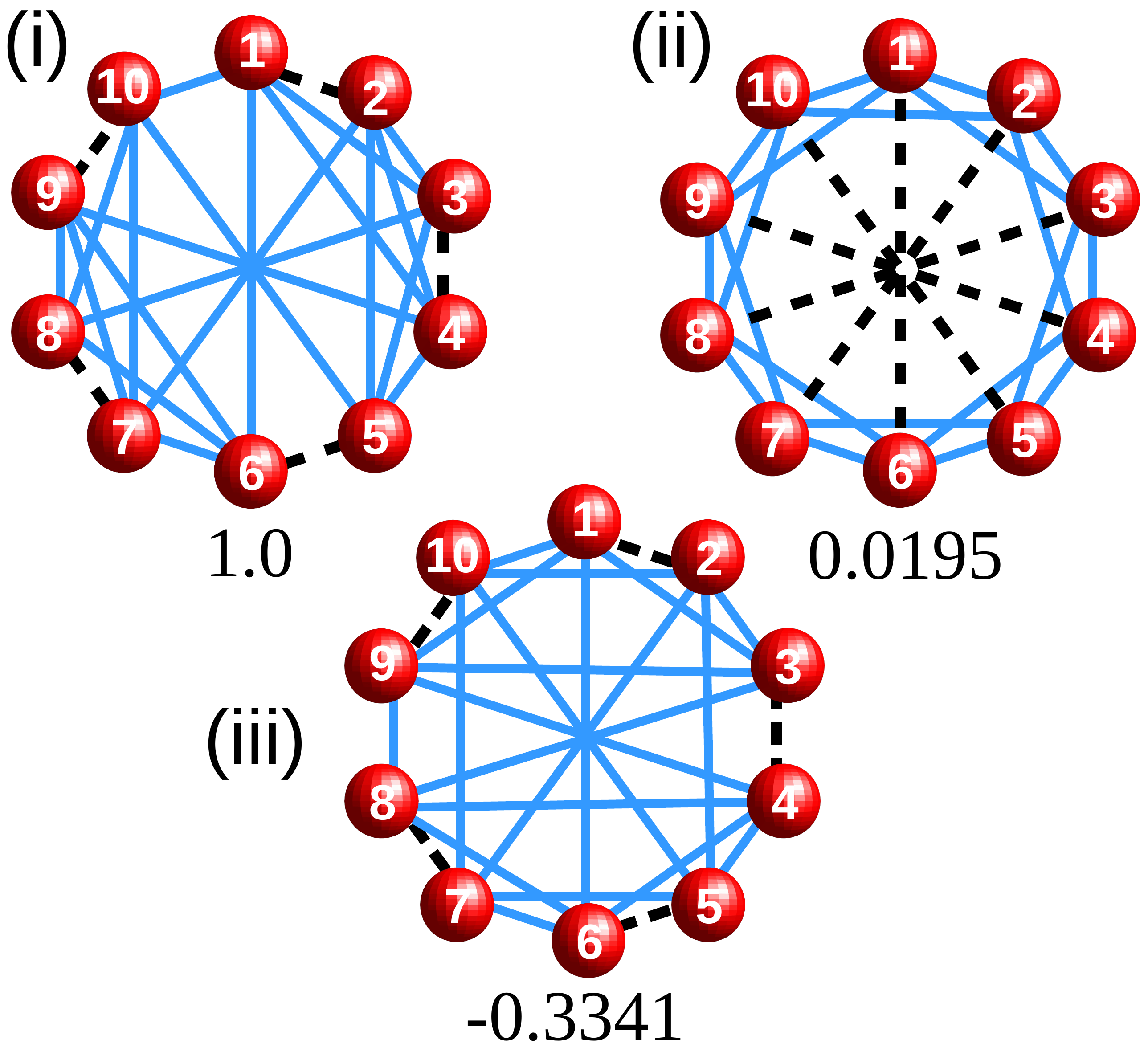}
		\end{minipage} 
		&\rotatebox[]{90}{ 0.9946(2)} \\
		\hline\hline
	\end{tabular}
\end{table}

The FPR for an incompressible $\nu =2/3$ state is \cite{Wu_93} $N_\phi = (3/2)N$ for even $N$. The morphology for $N=2$ is given by ${\cal M}_{2/3}^{(2)} = [3_1]$ as $N_\phi =3$.
For every increase of 2 electrons, number of flux quanta increases by 3 and thus two new entries in the set of morphology will be 2 and 1. Therefore, ${\cal M}_{2/3}^{(4)}
=[3_1,2_2,1_3]$, ${\cal M}_{2/3}^{(6)}=[3_1,2_2,2_3,1_4,1_5]$ etc. Factoring out the morphology corresponding to bosonic Laughlin state ${\cal M}_{L,1/2}^{(N)}
=[2_1,\cdots,2_{N-1}]$, the respective reduced morphologies become $\bar{{\cal M}}_{2/3}^{(4)} = [1_1,0_2,(-1)_3]$ and
$\bar{{\cal M}}_{2/3}^{(6)}=[1_1,0_2,0_3,(-1)_4,(-1)_5]$. Note that ${\cal M}_{2/3}^{(4)} = {\cal M}_{2/5}^{(4)}$ and hence $\bar{{\cal M}}_{2/3}^{(4)} 
=\bar{{\cal M}}_{2/5}^{(4)}$. 
Therefore, 
 $\Psi_{2/3}^{(4)}=\Psi_{2/5}^{(4)}$ which is expected as they both correspond to same number
of electrons and flux quanta. Three distinct graphs
 (Table-\ref{Table2}) can be constructed for $\bar{{\cal M}}_{2/3}^{(6)}$,
 out of which two produce lineraly indepenedent functions. A linear combination of the wave functions
obtained for these two graphs produces almost exact ground state wave function. Table-\ref{Table2} further shows the reduced morphologies for $N=8$ and $10$ electrons at $\nu = 2/3$ and the corresponding minimum number of possible graphs whose contributions lead to construct almost exact ground state wave functions.

%

\begin{table}[h]
	\caption{Same as described for $\nu=2/5$ in Table-\ref{Table1} replacing $\nu$ by $3/7$ and $3/5$.
	}
	\label{Table3}        
	\begin{tabular}{ c|c | c |c| c | c  } \hline\hline
		& & & & \\
		$\,\,N\,\,$& $\,\,2Q\,\,$ & $\bar{{\cal M }}_{3/7}^{(N)}$ & $\bar{{\cal M }}_{3/5}^{(N)}$ &  Graphs  &{$\,\mathcal{O}\,$}\\
		& & & & & \\ \hline\hline
		3  & 2  & \rotatebox[]{90}{ { \scriptsize $\big[(-1)_1,(-1)_2\big]$}}  &-- & \begin{minipage}{.25\textwidth}
			\includegraphics[width=13mm]{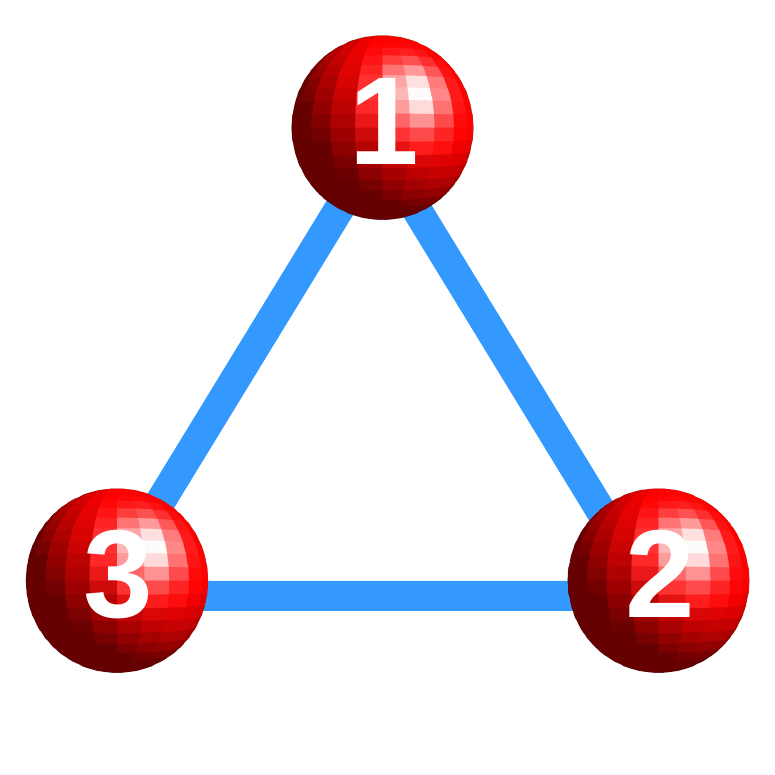} \end{minipage}   
		&\rotatebox[]{90}{1.0} \\
		  & 6  & --&\rotatebox[]{90}{{ \scriptsize $\big[1_1,1_2\big]$} }  & \begin{minipage}{.25\textwidth}
			\includegraphics[width=13mm]{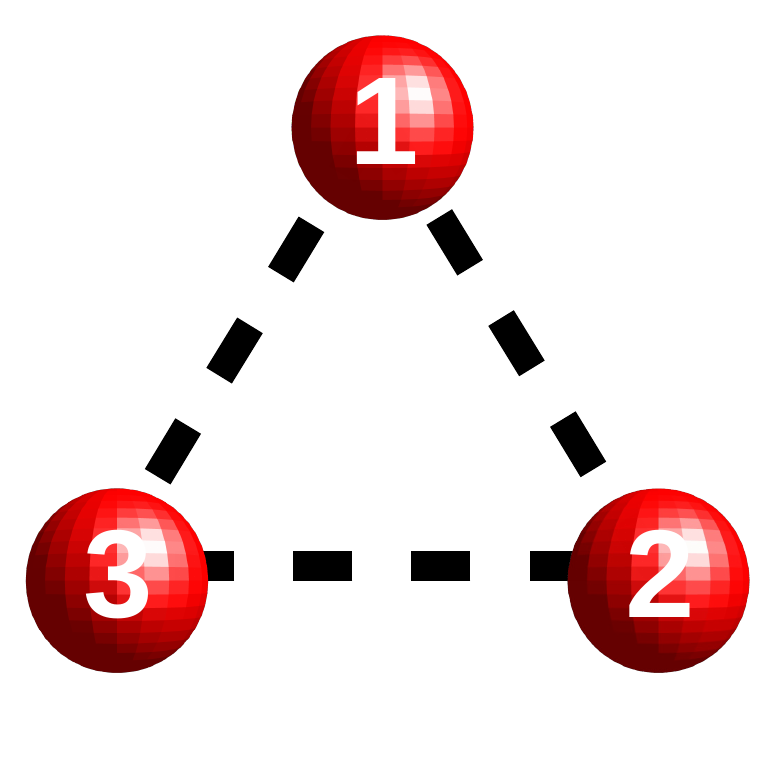} \end{minipage}   
		&\rotatebox[]{90}{1.0} 
		\\\hline
		6  & 9  &\rotatebox[]{90} {{ \scriptsize $\big[(-1)_1,(-1)_2,$$0_3,0_4,1_5\big]$}} &-- 
		&	\begin{minipage}{.25\textwidth}
			\includegraphics[width=30mm]{f23_N6.pdf}
		\end{minipage} 	
		& \rotatebox[]{90}{0.9999(0)} \\
			  & 11 &-- &\rotatebox[]{90}{ {\scriptsize $\big[1_1,1_2,0_3,0_4,(-1)_5\big]$}} &
			\begin{minipage}{.25\textwidth}
				\includegraphics[width=30mm]{f25_N6.pdf}
			\end{minipage} 	
			&\rotatebox[]{90}{ 0.9998(1)} \\
		
		\hline
		9  & 16  & \rotatebox[]{90}{{ \scriptsize $\big[(-1)_1,(-1)_2,0_3,0_4,0_5,0_6,1_7,1_8\big]$} }   &\rotatebox[]{90}{{\scriptsize $\big[1_1,1_2,0_3,0_4,0_5,0_6,(-1)_7,(-1)_8\big]$}}
		& 		\begin{minipage}{.25\textwidth}
			\includegraphics[width=45mm]{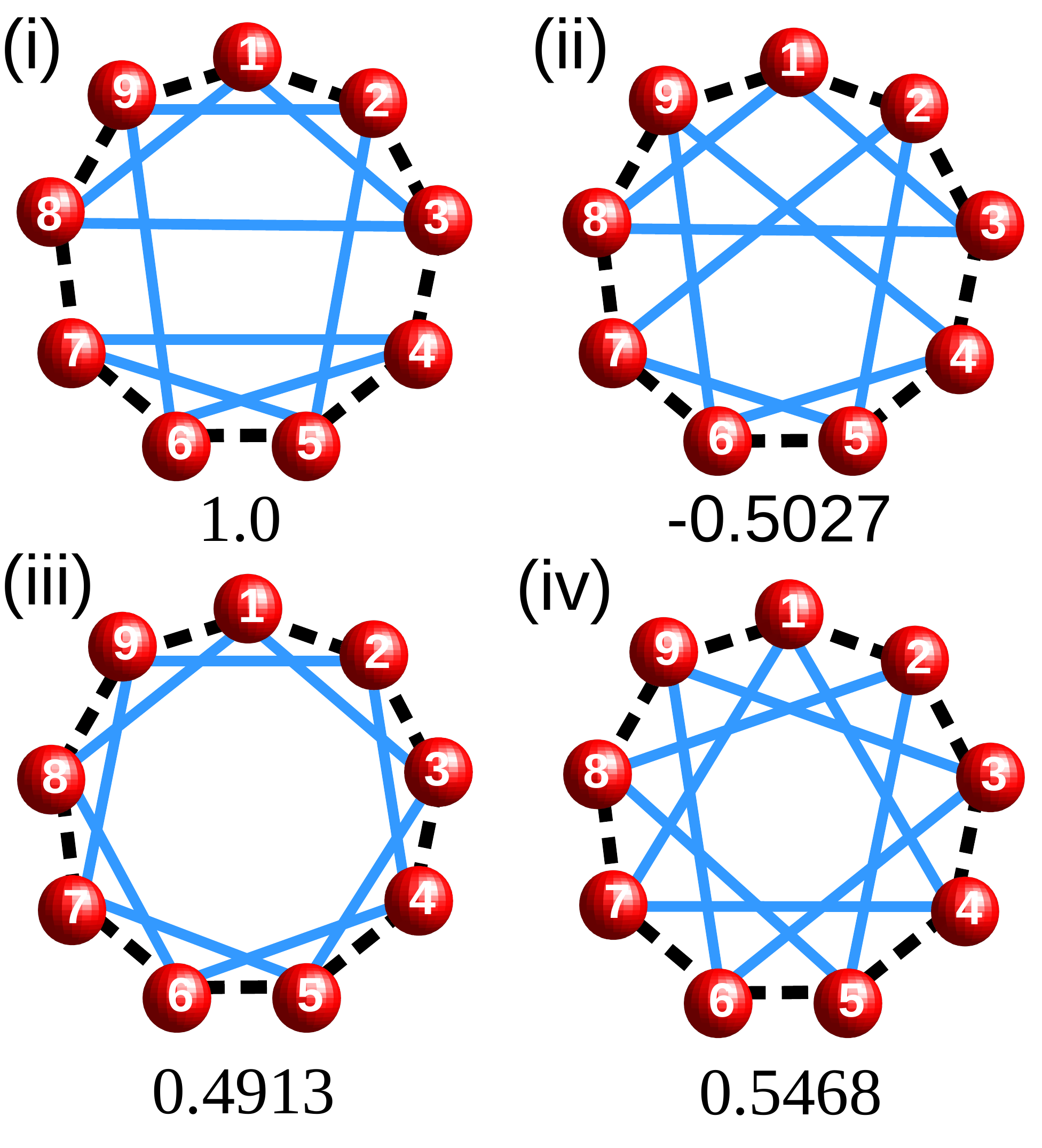} \end{minipage}  
		& \rotatebox[]{90}{0.9993(1)} \\ \hline
		\hline
	\end{tabular}
\end{table}

Table-\ref{Table3} shows reduced morphologies at $\nu = 3/7$ and $3/5$ for $N=3,\,6$ and $9$ with their respective FPR \cite{Wu_93} $N_{\phi}=(7/3)N-5$ and $N_{\phi}=(5/3)N+1$. Only one graph is possible for $N=3$ in each case and therefore the corresponding  morphologies construct the exact states, irrespective of the interaction. The wave functions constructed here is identical at $\nu = 3/7$ and $3/5$ for $N=9$ as they correspond to identical morphology that is expected due to the fact that both the states correspond to same number of flux quanta. We find that certain linear combination of the constructed wave functions from the graphs produce almost exact ground state of the Coulomb interaction.

In general, morphology for an $N$-electron system  at $\nu = 2/5$ will be ${\cal M}_{2/5}^{(N=2p)} = [1_1,2_2,2_3,\cdots , 2_{N/2},3_{N/2+1},\cdots , 3_{N-1}]$, where $p$ is an integer. The morphology for $\nu = 2/3$ whose FPR is $N_\phi = 3N/2$ is obtained as ${\cal M}_{2/3}^{(N=2p)} = [3_1,2_2,2_3,\cdots , 2_{N/2},1_{N/2+1},\cdots , 1_{N-1}]$. Similarly, the morphologies for $\nu = 3/7$ and $3/5$ are obtained respectively as ${\cal M}_{3/7}^{(N=3p)} = [1_1,1_2,2_3,\cdots , 2_{2N/3},3_{2N/3+1},\cdots , 3_{N-1}]$ and ${\cal M}_{3/5}^{(N=3p)} = [3_1,3_2,2_3,\cdots , 2_{2N/3},1_{2N/3+1},\cdots , 1_{N-1}]$. Based on these morphologies, the constructed wave-functions are shown to be almost exact up to N=10. We, thus, expect that these prescribed general morphologies will also be correct for any number of electrons. The exact ground state wave function may depend on the nature 
of the interaction, but the morphology of a given state will remain invariant; the actual wave-function will be the interaction-dependent different linear combinations 
of the antisymmetric functions obtained from the topologically distinct graphs of a particular morphology. Therefore, the morphology defines topology of a given FQHE state. We note that the overlaps are slighly less than 100\% due to the following reasons: (i) We have considered only two or three linearly independent graphs instead of taking all the possible linearly independent graphs for the same morphology; (ii) One morphology may not describe the exact ground state fully, for instance, the Laughlin wave function with a definite morphology is not \cite{MJ} an exact ground state for the Coulomb interaction. The tiny difference from the exact ground state wave function is for a negligible part due to other morphologies that can be obtained by the method of partitions \cite{MMR}.

  
  \begin{table}[h]
  \caption{Comparison of relative weight factors of the graphs tabulated in Tables \ref{Table1} and \ref{Table2} for
  	obtaining maximum overlap of the constructed wave functions \cite{Note} with the exact ground state of Haldane pseudo potential $V_1$ and Coulomb potential $V_C$.
  	Clearly, the ground state depends on the interaction potential, but topological structures remain universal.  }
    \label{Table4}
  	\begin{tabular}{ c| c| c |c |c |c | c | c}\hline\hline
  		$\,\nu\,$ & $\,N\,$ & $\,2Q\,$ & $~V~$ &\multicolumn{3}{|c|}{ Weight factors} & \,\,Overlap\,\, \\\cline{5-7}
  		 &  &  &  &$\quad(i)\quad$ &$(ii)$ &$(iii)$ &  \\
  		\hline\hline
  	
  		 & 6  & 11 & $V_c$ & 1.0 & -0.6928 & & $0.9998(1)$ \\
  		&    &    & $V_1$ & 1.0  &-0.7058&-- & $0.9994(1)$ \\\cline{2-8}
  	$2 \over 5$	& 8  & 16 & $V_c$ & 1.0  &-0.7733 &--& $0.9997(1)$ \\
  		&    &    & $V_1$ & 1.0 & -0.8107 &--& 0.9955(1) \\\cline{2-8}
  		& 10 & 21 & $V_c$ & 1.0 & 0.9259   & 0.5857& $0.9982(1)$ \\
  	
  		&    &    & $V_1$ & 1.0 & 0.8885 & 0.4866 & $0.9927(2)$ \\\hline
  		 & 6  & 9 & $V_c$ & 1.0 & 0.4304& --& $0.9999(0)$ \\
  		&    &    & $V_1$ & 1.0 & 0.3375&-- & $0.9999(0)$ \\\cline{2-8}
  		$2 \over 3$ & 8  & 12 & $V_c$ & 1.0 & 0.0343&-- & $0.9999(0)$ \\
  		&    &    & $V_1$ & 1.0 & 0.0494&-- & 0.9999(0) \\\cline{2-8}
  		& 10 & 15 & $V_c$ & 1.0 & 0.0195  & -0.3341 & $0.9946(2)$ \\

  		&    &    & $V_1$ & 1.0 &   0.0144 &-0.5162 & 0.9919(2) \\
  
  		\hline   
  		
  		\hline
  		
  	\end{tabular}	
  \end{table}   
  
 The topological structure of a given fractional quantum Hall effect is universal as we find that it depends on the graphs that can be constructed from one principle that every electron feels equal or unequal order of zeros at the positions of other electrons with the constraint that they all feel same set of order of zeros. Different linear combinations of the constructed wave functions from these graphs produce almost exact ground state wave functions for different interaction potentials. In Table-\ref{Table4}, we show a comparison of weight factors of different graphs for
 Coulomb interaction $V_C$ and Haldane-pseudopotential \cite{Haldane83} $V_1$ in the linear combination of the wave functions that give rise to almost exact ground state wave functions for both the interactions.

The morphologies of general filling factors $\nu = n/(2n+1)$ and $n/(2n-1)$ for same $n$ are closely related as they are interchangeable by the transformation $1_k \to 3_k$ and {\it vice versa}. As $n$ increases, the number of entries of $2$ in the set of morphology increases as $(n-1)[N/n-1]$ and the difference between the number of entries of $1$ and $3$ becomes $N/n-n$ when more number of $3$'s ($1$'s) occur in $\nu = n/[2n+1]$ ($n/[2n-1]$). Because of this symmetry in morphology, certain properties of these states are identical and the microscopic difference in morphology leads to certain different properties: Quasparticle charges are different while identical number of magneto-roton minima in their collective modes \cite{Girvin_PRL85,Scarola_PRB00,Majumder14}. Considering the symmetries of the entries in the set of morphology, the filling factor $\nu = 1/2$ through which these two sequences of filling factors are analytically continued for large $n$, will have equal number of $3$'s and $1$'s and that amounts to $\sqrt{N}-1$ numbers of $1$'s and $3$'s, and $N-2\sqrt{N}+1$ number of $2$'s as $N=n^2$ in this case. Indeed, the relation $N=n^2$ is consistent with the composite fermion picture in spherical geometry for studying $\nu = 1/2$ with finite number of electrons \cite{RR94}. If we generalize morphology from $\nu =n/(2n+1)$ to $\nu = n/(2sn+1)$, the elements in the set will renormalize as $3_k \to (2s+1)_k$,  $2_k \to (2s)_k$, and $1_k \to (2s-1)_k$. Further generalization for $\nu = n/(2sn -1)$ can be made by changing $ (2s+1)_k \to
(2s-1)_k$, $(2s-1)_k \to (2s+1)_k$ and keeping $(2s)_k$ fixed. Neutral excitation at angular momentum $L$ can be obtained by removing $L$ zeros for only one particle from a morphology, as previously has been studied \cite{Yang13} for Laughlin state.


\begin{references}
\bibitem{Note} A diagonalization of the interaction Hamiltonian in the subspace of the functions produced by the graphs will determine the ground state. Alternatively, one can consider exact ground state and then find the appropriate linear combination of these functions that provides maximum overlap with the exact state. We have performed the second procedure in this paper.
\end{references}
\end{document}